\documentclass[reprint, amsmath,amssymb,]{revtex4-2}
\usepackage{graphicx}
\usepackage{dcolumn}
\usepackage{bm}
\usepackage{hyperref}
\usepackage{color}
\usepackage{siunitx}
\usepackage{mathrsfs}
\usepackage{amsmath}
\makeatletter
\newcommand*\bigcdot{\mathpalette\bigcdot@{.7}}
\newcommand*\bigcdot@[2]{\mathbin{\vcenter{\hbox{\scalebox{#2}{$\m@th#1\bullet$}}}}}
\makeatother
\makeatletter
\usepackage[explicit]{titlesec}

\titlespacing*{\section}{0pt}{4.0ex plus .8ex minus 0.5ex}{1.6ex plus .0ex}
\titlespacing*{\subsection}{0pt}{3.5ex plus .0ex minus .0ex}{2.3ex plus .0ex}
\hypersetup{hypertex=true, colorlinks=true, linkcolor=blue, anchorcolor=blue, citecolor=blue}

\begin{document}
	
	\preprint{APS/123-QED}
	
	\title{Tetragonal Mexican-Hat Dispersion and Switchable Half-Metal State with Multiple Anisotropic Weyl Fermions in Penta-Graphene}
	
	\author{Ningning Jia\textsuperscript{1}, Yongting Shi\textsuperscript{1}, Zhiheng Lv\textsuperscript{1}, Junting Qin\textsuperscript{1},
		Jiangtao Cai\textsuperscript{2}, Xue Jiang\textsuperscript{3}, Jijun Zhao\textsuperscript{3}}
	\email{zhaojj@dlut.edu.cn(J.Zhao)}

    \author{Zhifeng Liu\textsuperscript{1}}
    \email{zfliu@imu.edu.cn(Z.Liu)}

	\affiliation{%
		$^1$Key Laboratory of Nanoscience and Nanotechnology, School of Physical Science and Technology, Inner Mongolia
		University, Hohhot 010021, China\\
		$^2$Department of Physics, Shaanxi University of Science and Technology, Xi’an
		710021, China\\
	    $^3$Key Laboratory of Materials Modification by Laser, Ion and Electron Beams (Dalian University of Technology), Ministry of Education, Dalian 116024, China}
    
\date{\today}
	
\begin{abstract}
	\noindent In past decades, the ever-expanding library of 2D carbon allotropes has yielded a broad range of exotic properties for the future carbon-based electronics. However, the known allotropes are all intrinsic nonmagnetic due to the paired valence electrons configuration. Based on the reported 2D carbon structure database and first-principles calculations, herein we demonstrate that inherent ferromagnetism can be obtained in the prominent allotrope, penta-graphene, which has an unique Mexican-hat valence band edge, giving rise to van Hove singularities and electronic instability. Induced by modest hole-doping, being achievable in electrolyte gate, the semiconducting penta-graphene can transform into different ferromagnetic half-metals with room temperature stability and switchable spin directions. In particular, multiple anisotropic Weyl states, including type-I and type-II Weyl cones and hybrid quasi Weyl nodal loop, can be found in a sizable energy window of spin-down half-metal under proper strains. These findings not only identify a promising carbon allotrope to obtain the inherent magnetism for carbon-based spintronic devices, but highlight the possibility to realize different Weyl states by combining the electronic and mechanical means as well.
\end{abstract}
\maketitle
\UseRawInputEncoding

\section{INTRODUCTION}
As the king of elements, carbon has abundant allotropes in different dimensionalities due to its bonding flexibility with itself \cite{RN4,RN18,RN5,RN22}. The best-known and well-studied carbon forms include three-dimensional (3D) graphite and diamond \cite{RN5}, two-dimensional (2D) graphene \cite{RN19}, one-dimensional (1D) nanotube \cite{RN20}, and zero-dimensional (0D) fullerene \cite{RN21}. In the past decades, the ever-expanding library of carbon allotropes has become more and more colorful. Naturally, the structural diversity gives rise to a broad range of exotic properties \cite{RN23,RN24,RN26,RN4216,RN7,RN25,RN27,RN28,RN12,RN9,RN8,RN10,RN29,RN30,RN3,RN13,RN16,RN14}, such as semiconductivity \cite{RN23,RN24,RN26,RN4216}, metallicity \cite{RN7,RN25,RN27,RN28}, superhardness \cite{RN12,RN9,RN8,RN10,RN29,RN30}, and topological states \cite{RN3,RN13,RN16,RN14}, which are anticipated to realize the desirable applications in microelectronic devices, Li/Na/K ion batteries, industrial cutting, and future topological quantum computation, respectively.

Owing to the paired valence electrons configuration, the known carbon allotropes are intrinsic nonmagnetic, which formidably hinders their applications in spintronics. To generate local magnetic moments, some schemes such as transition-metal doping \cite{RN34,RN35,RN38,RN39,RN40}, cutting into nanoribbon \cite{RN31,RN37}, and functionalization \cite{RN33} have been introduced in different carbon materials, especially in the nanostructures. Experimentally, however, precise control of these external modulation methods still remains challenging. In this context, it would be very meaningful to achieve inherent magnetism in carbon allotropes.

It has been proposed that magnetism can be viewed as an inherent property in 2D semiconductors with Mexican-hat bands \cite{RN44}. In this kind of materials, since the caused van Hove singularities can lead to electronic instability, the much desired ferromagnetism will be readily induced by modest carriers doping \cite{RN44,RN43,RN47,RN48}. As a consequence, such materials have a natural advantage to realize electrical tuning of the magnetic state \cite{RN44} via gate voltage. Then, an interesting question rises ------ whether the fascinating Mexican-hat bands exist in the pristine 2D carbon allotropes ? If so, what kind of exotic electronic states would emerge under carriers doping ?   

Bearing the feature of Mexican-hat bands in mind, herein we first screen the electronic bands of all the known 2D carbon allotropes in the ``2D carbon structure database" \cite{RN50}. The results show that the prominent penta-graphene \cite{RN24} is the only system exhibiting Mexican-hat dispersion in its frontier bands near the Fermi level. Using first-principles calculations, a Mexican hat-like dispersion with a van Hove singularity is identified in the valence band edge of penta-graphene. In view of the quadruple rotational symmetry, we propose a modified dispersion function by means of analytical method, which can well describe the Mexican-hat band. When penta-graphene is doped by holes with modest density, robust long-range ferromagnetic (FM) states with room temperature stability can be obtained, including spin-down half-metal, bipolar semiconductor, and spin-up half-metal. Remarkably, multiple anisotropic Weyl states (\emph{e.g.}, type-I and type-II Weyl cones, and quasi hybrid Weyl loop holding both type-I and type-II linear dispersions) are induced near the Fermi level of the spin-down half-metallic state by applying proper strain. With the advantage of electronic and mechanical tuning, the switchable half-metallicity and anisotropic Weyl fermions of penta-graphene hold promise for carbon-based spintronic devices.  

\begin{figure}[t]
	\includegraphics[width=0.95\linewidth]{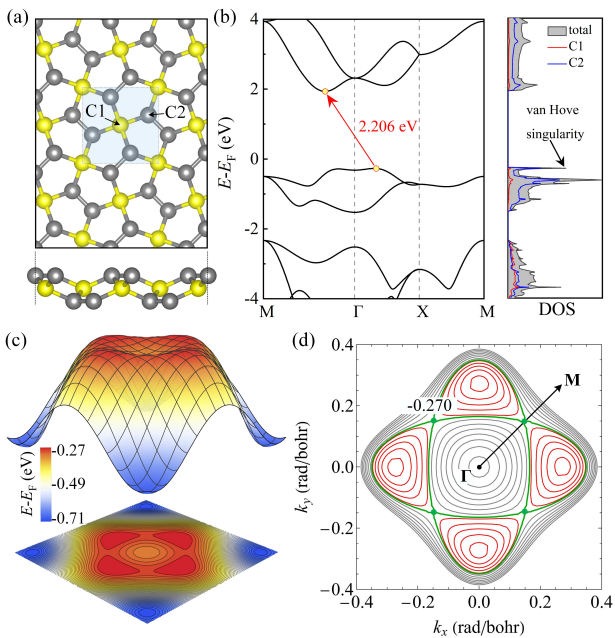}
	\caption{\label{fig:1}Structure and electronic bands of penta-graphene: (a) top and side view for the optimized structure, (b) electronic band structure and the corresponding density of state (DOS), (c) 3D plot together with its contours of the Mexican-hat valence band from DFT-PBE calculations, (d) energy contours of the fitted 3D Mexican-hat valence band using Eq. 2. For all the bands, the Fermi energy is set to 0 eV.}
\end{figure}

\section{COMPUTATIONAL METHODS}
We carry out state-of-the-art first-principles calculations based on density functional theory (DFT), as implemented with VASP \cite{RN3958}. The generalized gradient approximation (GGA) with Perdew-Burke-Ernzerhof (PBE) functional \cite{RN3959} is adopted to deal with the exchange-correlation interactions. For the description of ion-electron potential, the projector augmented wave (PAW) method \cite{RN4001} is employed by taking a cutoff energy of 600 eV for the plane-wave basis. A large vacuum layer (25 {\AA}) is added to the perpendicular direction of penta-graphene, so as to avoid the spurious interactions between the adjacent periodic images. A Monkhorst-Pack \textit{\textbf{k}}-point mesh with uniform spacing of $2{\pi}{\times}$0.02 {\AA}$^{-1}$ is used to sample the 2D Brillouin zone. The convergence criteria of $10^{-8}$ eV and 0.001 eV/{\AA} are used for the electronic iteration and Hellmann-Feynman force, respectively. 
\\\hspace*{\fill} \\                                                             
\section{RESULTS AND DISCUSSION}

\subsection{Mexican-hat dispersion of the valence band edge}
Like the Cairo pentagonal tiling \cite{RN65}, penta-graphene is made up entirely of carbon pentagons [see Fig. \ref{fig:1}(a)]. Its atomic structure holds \emph{P}$\overline{4}$$2_1$\emph{m} ($D_{2d}^{3})$ symmetry in a tetragonal lattice. The optimized lattice constant is a = b = 3.637 {\AA}, close to previously reported value of 3.64 {\AA} \cite{RN24}. In every unit cell, there are six carbon atoms that can be classified into two categories: C1 (yellow) and C2 (grey), holding four-coordinated $\emph{sp}^{3}$ and three-coordinated $\emph{sp}^{2}$ bonding environment, respectively. The electronic band structure plotted in Fig. \ref{fig:1}(b) reveals that penta-graphene is an indirect semiconductor with a bandgap of 2.21 eV at PBE level [3.28 eV using HSE06 functional (see Fig. S1 in the Supplemental Material \cite{S}, being well consistent with 3.25 eV in Ref \cite{RN24})]. Interestingly, the valence band edge of penta-graphene has a small local minimum at the $\Gamma$ point, which makes it shaped as a shallow inverted Mexican-hat [see Fig. \ref{fig:1}(c)]. Such remarkable band feature  corresponds to a sharp van Hove singularity divergence [Fig. \ref{fig:1}(b)] with 1$\slash$$\sqrt{E}$ on the density of state (DOS) \cite{RN44,RN47,RN48}. From the partial DOS, one can find that the Mexican-hat band primarily originates from C2 atoms.

To provide a quantitative description for the Mexican-hat band of penta-graphene, we propose a modified dispersion function ``$\emph{E}_v$(\textbf{\emph{k}})" similar to the case of hexagonal GaS monolayer \cite{RN49,RN46}. Taking angular dependence (quadruple rotational symmetry) into account, the dispersion function is written as 
\begin{equation}\label{func_1}
	E_v(\emph{k},\theta)=\mathrm{E}_{\mathrm{0}}+{\rm{A}}k^2+{\rm{B}}k^4+{\rm{F}}k^4 {\rm cos}(\theta)
\end{equation}
\noindent in the 2D polar-coordinate system, or expressed as 

\begin{widetext}
	\begin{equation}\label{func_2}
		E_v(k_x,k_y)=\mathrm{E}_{\mathrm{0}}+{\rm{F}}(k_x^{4}+k_y^{4})-6{\rm{F}}k_x^{2}k_y^{2}
		+{\rm{A}}(k_x^{2}+k_y^{2})+{\rm{B}}(k_x^{2}+k_y^{2})^{2}
	\end{equation}
\end{widetext}
in the Cartesian coordinate system. Thereinto, \emph{k} = $\lvert$\textbf{\emph{k}}$\lvert$ = ($\emph{k}_x^{2}$ + $\emph{k}_y^{2}$)$^{1/2}$. Based on the numerical DFT calculated valence band surface [Fig. \ref{fig:1}(c)], the dispersion function (2) is fitted in a square area of \textit{\textbf{k}} space centered at $\Gamma$ point with the side length of 0.4 rad/bohr. The values of $\mathrm{E}_{\rm{0}}$, A, B and F are confirmed as $-$0.28944 eV, 0.88301 eV ${\rm \AA^{2}}$, $-$7.90467 eV ${\rm \AA^{4}}$, and 2.00059 eV ${\rm \AA^{4}}$, respectively. Fig. \ref{fig:1}(d) depicts the contour plot of the Mexican-hat band obtained from the dispersion function (2) with the fitted $\mathrm{E}_{\rm{0}}$, A, B and F parameters. Comparing with the contour plot in Fig. \ref{fig:1}(c), one can see that the proposed dispersion function $\emph{E}_v$(\textbf{\emph{k}}) reproduces the DFT data well. Furthermore, our detailed analysis reveals that there is a saddle point ($\emph{k}_x$ = 0.145 $\rm \AA^{-1}$, $\emph{k}_y$ = 0.145 $\rm \AA^{-1}$, $\emph{E}_v$ = $-$0.270 eV) in the valence band along the $\Gamma$ $\to$ M (\emph{i.e.}, $\theta$ = $45^{\circ}$ direction) in the \textit{\textbf{k}}-space. Owing to its presence, the aforementioned 1D van Hove singularity in the density of states emerges. Moreover, a Lifshitz transition \cite{RN49,RN46} would occur when the Fermi level is modulated to cross the critical energy of the saddle point. This is due to the change of topology of the Fermi surface, \emph{i.e.}, from four isolated small rings, to interconnected rings [see the green lines in Fig. \ref{fig:1}(d)], then to one big ring.

\subsection{Magnetic properties and switchable half-metal states}
\begin{figure}[t]
	\includegraphics[width=1\linewidth]{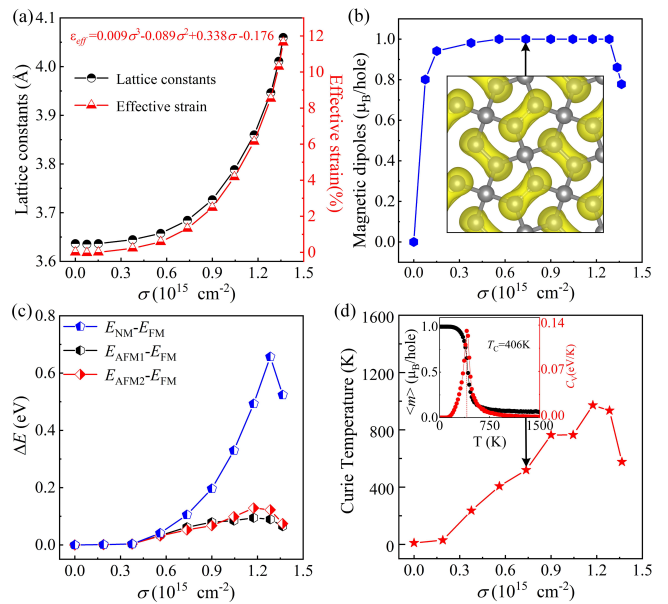}
	\caption{\label{fig:2}(a) Optimized lattice constants and the corresponding effective strains, (b) magnetic moment per hole, (c) relative energies to FM state, (d) Curie temperature of penta-graphene as a function of the doping densities $\sigma$. The inset in (b) is the spatial distribution of spin-polarized electron density at the hole density of 7.36$\times$$10^{14}$ ${\rm cm}^{-2}$, while that in (d) displays the temperature-dependent magnetic moments and heat capacity ($\emph{C}_{\rm {v}}$) for the penta-graphene at the hole density of 7.36$\times$$10^{14}$ ${\rm cm}^{-2}$.}
\end{figure}

Apart from the pristine penta-graphene, it would be more interesting to explore penta-graphene with \emph{p}-type doping. As mentioned above, 1D van Hove singularity is an important indication of electronic instabilities, providing the opportunity to generate different phases \cite{RN44,RN47,RN48,RN3660} such as ferromagnetism and superconductivity. Indeed, previous experiments have revealed that high doping level of electron or hole carriers up to $10^{15}$ ${\rm cm}^{-2}$ can be achieved in 2D materials with an electrolyte gate \cite{RN53,RN54,RN55}. For these, we first assess the impact of hole doping on the penta-graphene. With different holes densities, the structures of \emph{p}-type penta-graphene have been fully re-optimized using spin-polarized DFT calculations. The results show that C2 atoms tend to form a more standard planar $\emph{sp}^{2}$ hybridization with the increasing of densities, which enlarges the $\angle$C1C2C1 angle and lattice constants [see Fig. \ref{fig:2}(a)]. Indeed, such effect is equivalent to applying an effective biaxial tensile strain $\varepsilon_{eff}$ (up to 11.8$\%$) to the penta-graphene. The relationship between $\varepsilon_{eff}$ and hole density can be well characterized by a third-order polynomial fitting as: $\varepsilon_{eff}$ = 0.009$\sigma^3$$-$0.089$\sigma^2$$+$0.338$\sigma$$-$0.176. 

In addition to doping-induced structural changes, increasing hole doping also breaks the time-reversal symmetry and leads to FM state in penta-graphene. As plotted in Fig. \ref{fig:2}(b), the localized magnetic moment of per hole reaches up to 1 $\mu_{\rm B}$/hole at the holes densities ranging from 5.60$\times$$10^{14}$ ${\rm cm}^{-2}$ to 1.28$\times$$10^{15}$ ${\rm cm}^{-2}$. The corresponding spin charge density reveals that all magnetic moments are mainly contributed by the C2-C2 dimers, which can be viewed as one magnetic moment unit [see inset of Fig. \ref{fig:2}(b)]. Because of this, we construct a 2 $\times$ 2 supercell, and then calculate their total energy in different collinear magnetic orders, including FM state, possible antiferromagnetic states (AFM1 and AFM2, see Fig. S2 in the Supplemental Material \cite{S}), and nonmagnetic state NM. The relative energies of AFM1, AFM2 and NM with respect to FM state are displayed in Fig. \ref{fig:2}(c). Evidently, FM order is always the most stable magnetic state for the \emph{p}-type penta-graphene for all the considered holes densities. 

After confirming the magnetic ground state, it is highly necessary to examine its thermal stability, which is essential for practical spintronic applications. For this, we estimate the Curie temperature ($T{\rm_C}$) of \emph{p}-type FM penta-graphene with different hole densities. By considering the nearest-neighbor ($J{\rm_1}$) and next-neighbor ($J{\rm_2}$) exchange parameters, the Hamiltonian of penta-graphene in a classical Heisenberg model can be expressed as 
\begin{equation}\label{func_3}
	\textit{H}=-\sum_{i, j} \emph{J}_{1} M_{i} M_{j}-\sum_{k, l} J_{2} M_{k} M_{l},
\end{equation}
Here, \textit{M}$_{x}$ (\textit{x} = \textit{i}, \textit{j}, \textit{k}, \textit{l} ) denotes the magnetic moment on different sites, denoting every C2-C2 dimer in magnetic penta-graphene. Thus, from the total energies of different magnetic states, $J{\rm_1}$ and $J{\rm_2}$ parameters can be evaluated by 
\begin{equation}\label{func_4}
	J_{1}=\frac{E(\text { AFM1 })-E(\mathrm{FM})}{8 M^{2}},
\end{equation}
\begin{equation}\label{func_5}
	J_{2}=\frac{E(\mathrm{FM})+E(\mathrm{AFM} 1)-2E(\mathrm{AFM} 2)}{-8 M^{2}}.
\end{equation}

\noindent Based on Wolff algorithm \cite{RN57}, Monte Carlo simulations are carried out on a 16 $\times$ 16 lattice by using the Mcsolver code \cite{RN56}. Overall, the obtained $T{\rm_C}$ increases with the increasing of holes density [see Fig. \ref{fig:2}(d)] due to the enhance exchange energy [\emph{$\Delta$E}$_{\rm FM}$ = \emph{E}$_{\rm NM}$$-$\emph{E}$_{\rm FM}$, see Fig. \ref{fig:2}(c)]. Specifically, (i) $T{\rm_C}$ first increases to the maximum of 972 K at 1.28$\times$$10^{15}$ ${\rm cm}^{-2}$ and then decreases, consistent with the tendency of relative energies of different magnetic states [see Fig. \ref{fig:2}(c)]; (ii) all the $T{\rm_C}$ values for the hole densities ranging from 7.36$\times$$10^{14}$ ${\rm cm}^{-2}$ to 1.37$\times$$10^{15}$ ${\rm cm}^{-2}$ are larger than room temperature (see Fig. S3 in the Supplemental Material \cite{S}), ensuring that the FM order is robust enough for applications in the ambient environment.

\begin{figure}[t]
	\includegraphics[width=1.\linewidth]{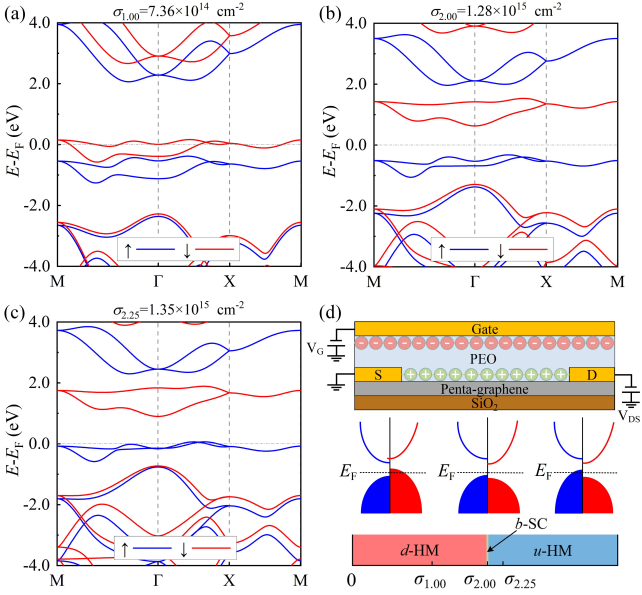}
	\caption{\label{fig:3}Electronic band structures of penta-graphene with hole doping at different densities: (a) 7.36$\times$$10^{14}$ ${\rm cm}^{-2}$ with one hole per primitive cell, (b) 1.28$\times$$10^{15}$ ${\rm cm}^{-2}$ with two hole per primitive cell, and (c) 1.35$\times$$10^{15}$ ${\rm cm}^{-2}$ with 2.25 hole per primitive cell. (d) Schematic diagrams of electrolytic gate and band structures for different spin-polarized states: \emph{d}-HM, \emph{b}-SC and \emph{u}-HM.}
\end{figure}

Besides the robust FM order, three appealing electronic states with distinct conductive properties are observed in the calculated band structures of \emph{p}-type penta-graphene [see Fig. S4 in the Supplemental Material \cite{S}], including spin-down half-metal (\emph{d}-HM), bipolar semiconductor (\emph{b}-SC), and spin-up half-metal (\emph{u}-HM). Among them, 100$\%$ spin-polarized \emph{d}-HM state is formed at hole densities from 1.9$\times$$10^{14}$ ${\rm cm}^{-2}$ to 1.17$\times$$10^{15}$ ${\rm cm}^{-2}$, in which the transport carriers are purely spin-down holes. Fig. \ref{fig:3}(a) presents a typical example of \emph{d}-HM at the density of 7.36$\times$$10^{14}$ ${\rm cm}^{-2}$. One can see that there is a half-filled spin-down band across the Fermi energy. When the density increases to 1.28$\times$$10^{15}$ ${\rm cm}^{-2}$, penta-graphene transforms into a non-conducting \emph{b}-SC [Fig. \ref{fig:3}(b)], in which the conduction and valence bands have opposite spin directions, exhibiting bipolar characteristics \cite{RN67}. Since there is no conductive carrier, such \emph{b}-SC should be regarded as an indispensable off-state in device application. As for the larger density of 1.37$\times$$10^{15}$ ${\rm cm}^{-2}$, penta-graphene is switched to conductive half-metal again. However, the transport carriers become spin-up holes instead of spin-down ones, corresponding to the \emph{u}-HM state, as shown in Fig. \ref{fig:3}(c). In experiments, as hole density can be tuned by the electrolytic gate \cite{RN53,RN54,RN55} [see schematic diagram in Fig. \ref{fig:3}(d)], \emph{p}-type penta-graphene should be an ideal platform to achieve switchable and reversible spintronic device, in which the transformation among \emph{d}-HM $\leftrightarrow$ \emph{b}-SC $\leftrightarrow$ \emph{u}-HM states can be readily realized by electrical means, as illustrated in Fig. \ref{fig:3}(d). 

\subsection{Strain induced multiple Weyl states}

\begin{figure*}[t]
	\includegraphics[width=0.65\linewidth]{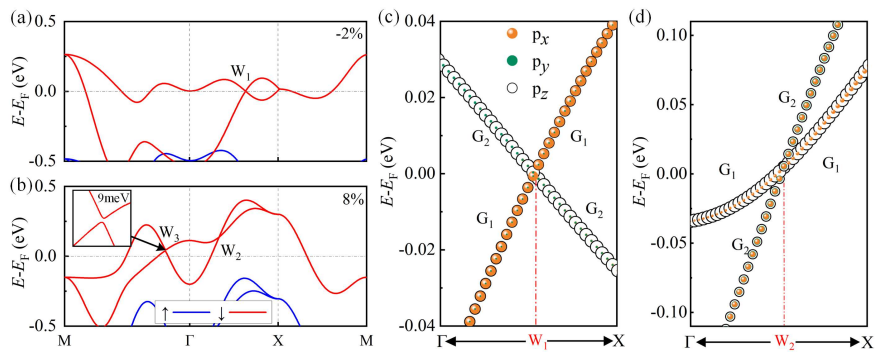}
	\caption{\label{fig:4}(a) Electronic band structures of penta-graphene at the density of 7.36$\times$$10^{14}$ ${\rm cm}^{-2}$ under compress strain of (a) $-$2$\%$ and tensile strain of (b) 8$\%$. The orbital-decomposed linear crossing bands with irreducible representations around (c) ${\rm W_1}$ and (d) ${\rm W_2}$.}
\end{figure*}

\begin{figure}[t]
	\includegraphics[width=1\linewidth]{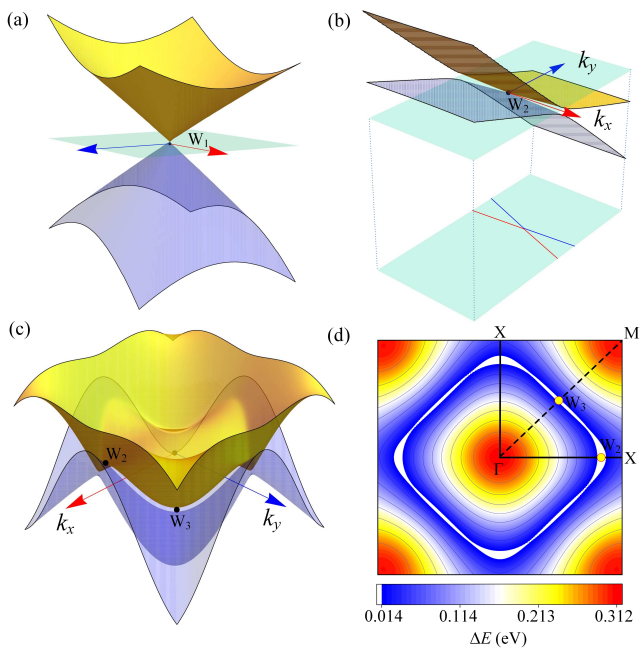}
	\caption{\label{fig:5}The 3D band structures on a defined \textit{\textbf{k}}-plane centered at (a) ${\rm W_1}$ under $-$2$\%$ compress strain and (b) ${\rm W_2}$ at 8$\%$ tensile strain, and (c) that on the whole 2D Brillouin zone with 8$\%$ strain. (d) The colorized contour plot of the energy gap between the two bands in (c). The cyan planes in (a)-(c) denote the energy isosurface of the corresponding Weyl point.}
\end{figure}

According to the dispersion feature of conducting bands around the Fermi level, there exists linear crossing along the $\Gamma$ $\to$ X path in the spin-down channel of penta-graphene at hole densities of 5.6$\times$$10^{14}$ ${\rm cm}^{-2}$ $\sim$ 9$\times$$10^{14}$ ${\rm cm}^{-2}$ (see Fig. S4 in the Supplemental Material \cite{S}). However, the effective energy window with linear bands is rather small, which will hinder the experimental detection of the corresponding massless fermions. Then, an interesting question arises: can the linear energy window and the band dispersion be modulated by a feasibly external method ? To explore this, taking the penta-graphene at 7.36$\times$$10^{14}$ ${\rm cm}^{-2}$ as an example, we re-calculate its band structures under different mechanical strains, which has been confirmed as an accessible approach for engineering the electronic properties of 2D materials \cite{RN58,RN59}. As seen from the band structures in Fig. S5 in the Supplemental Material \cite{S}, the crossing bands are sensitive to both the tensile and compressive strains. Hereafter, we focus on the following two cases, \emph{i.e.}, at the compressive strain of $-$2$\%$ and at tensile strain of 8$\%$. 

Under the $-$2$\%$ strain, a double degenerate Weyl state [Fig. \ref{fig:4}(a)] with a sizable linear energy window ($\sim$140 meV) can be found along the $\Gamma$ $\to$ X path at the Weyl point of ${\rm W_1}$ (0.30936, 0.00000). The computed 3D band structure reveals that this Weyl point is isolated in the 2D Brillouin zone, and the two low-energy bands around the Fermi level form a tetragonal symmetric type-I Weyl cone [Fig. \ref{fig:5}(a)] with weak transport anisotropy, whose Fermi velocities along different \textit{\textbf{k}} directions range from 1.55$\times$$10^{5}$ m$\slash$s to 1.90$\times$$10^{5}$ m$\slash$s (see Fig. S6(b) in the Supplemental Material \cite{S}). To explore the origin of such Weyl state, we further analyze the orbital-resolved bands. As displayed in Fig. \ref{fig:4}(c), the wavefunctions around ${\rm W_1}$ are mainly contributed by the \emph{p$_x$} and \emph{p$_z$} orbitals, and there is an obvious inversion of energy order between the \emph{p$_x$} and \emph{p$_z$} bands. Such band inversion is one of the important signals of topological semimetals \cite{RN68,RN69,RN70}. Additionally, to elucidate how the Weyl state forms, we perform symmetry analysis for the two crossing bands using the IRVSP code \cite{RN71}. Around ${\rm W_1}$, the obtained irreducible representations of the two linear bands in the high symmetry path are G1 and G2 [Fig. \ref{fig:4}(c)], respectively. Using Bilbao Crystallographic Server \cite{RN72}, it is found that the corresponding symmetrical element is dual rotational axis $\mathrm{C}_2$. Therefore, it is reasonable to conclude that the Weyl state around ${\rm W_1}$ is formed by the band inversions between \emph{p$_x$} and \emph{p$_z$} of carbon atoms with the protection of $\mathrm{C}_2$ rotational symmetry.

Interestingly, when 8$\%$ tensile strain is applied to penta-graphene, the original type-I Weyl cone evolves into a tilted type-II Weyl cone, in which the slopes of the two crossing bands in some \textit{\textbf{k}}-paths share the same sign around ${\rm W_2}$ [see Fig. \ref{fig:4}(b)]. This stems from that the kinetic component $\emph{T}$(\textbf{\emph{k}}) is larger than the potential component $\emph{U}$(\textbf{\emph{k}}) in the energy spectrum of \emph{E}$_{\pm}$(\textbf{\emph{k}}) = $\emph{T}$(\textbf{\emph{k}}) $\pm$ $\emph{U}$(\textbf{\emph{k}}) for a Weyl state \cite{RN60}. A more visualized image can be seen in the 3D bands with a small \textit{\textbf{k}}-plane centered at ${\rm W_2}$ point, as shown in Fig. \ref{fig:5}(b). There  coexist electron and hole pockets on the energy isosurface of ${\rm W_2}$, as presented in the lower inset. Moreover, band dispersion of the Weyl cone, unlike the case of Fig. \ref{fig:5}(a), is highly anisotropic (see Fig. S6(c) in the Supplemental Material \cite{S}). Along the \textbf{$\emph{k}_x$} direction, the Fermi velocity reaches up to a maximum of 2.7$\times$$10^{5}$ m$\slash$s, while in the $\theta$ = $72.74^{\circ}$ direction, the Fermi velocity is reduced to a minimum value of 2.5$\times$$10^{3}$ m$\slash$s due to the weak band dispersion. 

Besides ${\rm W_2}$, another Weyl crossing point, \emph{i.e.}, ${\rm W_3}$ (0.16638, 0.0000) along $\Gamma$ $\to$ M, also emerges under 8$\%$ tensile strain. Unlike ${\rm W_2}$, however, it is not a strict degeneracy point because of the existence of a 9 meV tiny bandgap [see inset in Fig. \ref{fig:4}(b)]. This can be understood by the mechanism of band repulsion \cite{RN62}, namely, when there is no symmetry (see Fig. S7 in the Supplemental Material \cite{S}), the two crossing bands will hybridize with each other and maintain a gap in-between. Even so, considering the thermal effect in practical environment, such a tiny bandgap should be ignored at temperatures higher than 100 K. In this regard, ${\rm W_3}$ can be considered as a massive Weyl point. Then, multi-crossing Weyl points are reminiscent of the FM Weyl nodal lines \cite{RN63,RN64}, which is composed of consecutive Weyl points. To explore the complete pattern of the Weyl points in the 2D Brillouin zone, we plot the corresponding 3D band structure. As presented in Fig. \ref{fig:5}(c), there are four equivalent and strict Weyl points protected by $\mathrm{C}_2$ symmetry, and infinite number of massive Weyl points, like ${\rm W_3}$. The colorized contour plot [Fig. \ref{fig:5}(d)] of the energy gap between the two bands further reveals that all the ``Weyl points" form a closed quasi nodal line shaped as a rounded square centered at point $\Gamma$. Moreover, in the view of band dispersion, the quasi nodal line should be referred to as a hybrid nodal loop ------ there exist both type-I and tilted type-II Weyl dispersions along the transverse directions of the nodal line, as displayed in the bands along $\Gamma$ $\to$ M and $\Gamma$ $\to$ X, respectively. As proposed, such type of hybrid nodal loop may lead to unconventional magnetic responses \cite{RN3662}, such as the effect of zero-field magnetic breakdown and the peculiar anisotropy in the cyclotron resonance.

\section{CONCLUSIONS}
Using first-principles calculations, we show that the prominent penta-graphene is the only system harboring the novel Mexican-hat band edge among all known 2D carbon allotropes. An analytical dispersion function with quadruple rotational symmetry is proposed to describe the Mexican-hat band. Owing to the existence of van Hove singularities and electronic instability, the desirable room temperature FM half-metal states with switchable spin directions can be readily obtained by achievable hole doping. Remarkably, both weak and strong anisotropic Weyl states, including type-I/type-II Weyl cones and quasi hybrid Weyl nodal loop, are found in the spin-down half-metal state under modest strains. In view of the accessibility of high-density carrier doping by electronic means, we believe that the robust FM half-metal states identified in penta-graphene would be promising in carbon-based spintronic devices. Additionally, our findings also highlight the strategy to obtain different Weyl fermions by combining the electronic and mechanical modulations.

\section*{ACKNOWLEDGMENTS}
This work is supported by the National Natural Science Foundation of China (11964023, 91961204, 11874097,
12274050), Natural Science Foundation of Inner Mongolia Autonomous Region (2021JQ-001), and the 2020 Institutional Support Program for Youth Science and Technology Talents in Inner Mongolia Autonomous Region (NJYT-20-B02).
~\\
\bibliography{C6-manuscript}

\end{document}